\documentclass[prb,aps,showpacs,twocolumn,unsortedaddress]{revtex4}
\usepackage{graphics,bm}
\usepackage{amssymb}
\usepackage{epsfig}
\usepackage{epsf}
\begin{document}

\title{Modulation Doping near Mott-Insulator Heterojunctions}

\author{Wei-Cheng Lee}
\email{leewc@mail.utexas.edu}

\author{A.H. MacDonald}
\email{macd@ph.utexas.edu}
\affiliation{Department of Physics, The University of Texas at Austin, Austin, TX 78712}

\date{\today}

\begin{abstract}
We argue that interesting strongly correlated two-dimensional electron systems 
can be created by modulation doping near a heterojunction between Mott 
insulators.  Because the dopant atoms are remote from
the carrier system, the electronic system will be weakly disordered.
We argue that the competition between different ordered states can
be engineered by choosing appropriate values for the dopant density and 
the setback distance of the doping layer.  In particular larger setback
distances favor two-dimensional antiferromagnetism over ferromagnetism.
We estimate some key properties of modulation-doped 
Mott insulator heterojunctions by combining insights from Hartree-Fock-Theory and 
Dynamical-Mean-Field-Theory descriptions and discuss potentially
attractive material combinations.  
\end{abstract}

\maketitle

\section {Introduction}
The electronic properties of transition metal oxides\cite{imada,tokura,dagotto} are determined by
delicate balancing acts involving hybridized oxygen {\em p} and transition-metal {\em d}
orbitals and strong correlations that suppress charge fluctuations on
transition metal sites.  Magnetic and transport properties in these materials
are extraordinarily sensitive 
to the character of the orbitals present at the Fermi energy, and therefore
to external influences like doping and strain.  This sensitivity has motivated
interest in the epitaxial growth of oxide heterojunctions and
artificial layered oxides. This line of research seeks im part to emulate the 
achievements of semiconductor materials researchers who have over the past few decades 
learned to engineer the electronic properties of epitaxially grown semiconductor materials 
by exploiting lattice-matching strains and modulation doping.
Because of the greater sensitivity of their electronic 
properties and because of the wider range of phenomena (particularly magnetic 
phenomena) that occur, the implications for physics and for technology of 
substantial advances in the oxide case are likely to 
be enormous.  In anticipation of future progress on the materials side, 
we explore in this paper some of the physics of modulation doping 
in epitaxially grown transition metal oxides, emphasizing differences between the 
strongly-correlated-material case and the familiar semiconductor heterojunction case.
We find that strong correlations enhance the two-dimensional character of 
the metals that occur near modulation-doped heterojunctions and
the range of doping which is possible without producing unwanted parallel
conduction.  We argue that two-dimensional electron systems produced in this
way are likely to have remarkable properties and that modulation doping near interfaces 
between two-different Mott insulators may make it 
possible not only to create weakly-disordered low-dimensional strongly correlated 
electron systems, but also to engineer the compromises that occur in these 
systems between different types of magnetic order.  

Transition metal oxides are prototypical strongly-correlated-electron systems.
Understanding their electronic properties has been one of the most challenging topics 
in condensed matter theory.  Single-particle energy scales like the widths
of bands near the Fermi energy are  
often comparable to or smaller than characteristic interaction energy scales, 
challenging band theory descriptions.  Several different classes of 
transition metal oxides have been studied extensively
revealing various interesting types of order
involving spin, charge, and orbital degree of freedoms\cite{imada,tokura,dagotto}
and leading to fundamental discoveries like high-$T_c$ superconductivity and colossal magnetoresistance.
In the last decade, notable progress has been made in manipulating transition metal oxides
by gating and by controlled layer by layer growth. Ahn {\it et al.} have
applied the field-effect approach to ferroelectric oxide/high-T$_c$ cuprate
heterostructures and successfully tuned superconducting properties.\cite{ahn1,ahn2} 
Ohtomo {\it et al.}\cite{ohtomo} have observed unusual metallic 
behavior at Mott-insulator-band-insulator
(MIBI) heterostructures realized by precisely controlled growth of LaTiO$_3$/SrTiO$_3$ layers.
Very recently, Chakhalian\cite{keimer} {\em et al.} have studied the interplay between magnetic and superconducting 
order at an interface between (La,Ca)MnO$_3$ and YBCO. 
These achievements not only provide new platforms for fundamental
research, but also demonstrate the promise of devices with functionality that
is based on the unique properties of strongly correlated oxide materials. 

There have also been important advances in the theoretical description of artificially
layered transition metal compounds. Efforts have been made\cite{mp1,RM,ZF,mp2,AL} to understand
differences between surface and bulk properties in strongly-correlated
materials, providing insights into the main consequences of the absence of translational 
invariance along certain directions. First principles calculations for 
PbTiO$_3$/SrTiO$_3$ superlattices demonstrated interesting ferroelectric properties.\cite{rabe}
The effect of spatially inhomogeneous multilayered structures on transport properties has been examined.\cite{freericks}
Recently Okamoto and Millis\cite{okamoto1,okamoto2,okamoto3}
used combined insights from Hartree-Fock theory (HFT) and dynamical mean-field theory
(DMFT) to investigate the LaTiO$_3$/SrTiO$_3$ model MIBI heterostructure systems mentioned above,
and successfully described the interplay between long-range Coulomb interactions and 
strong short range correlations in the electron density distribution
near MIBI heterojunctions. They concluded that the unusual metallic behavior observed by Ohtomo 
{\it et al.} originates at the MIBI interface and that the properties of this 
interface are very different from those in the bulk, because of an {\em electronic surface reconstruction}
reminiscent of the purely electronic\cite{Sawatzky} reconstructions imposed by space-charge 
physics on systems with polar surface terminations\cite{Noguera2000}. 

\begin{figure}
\includegraphics{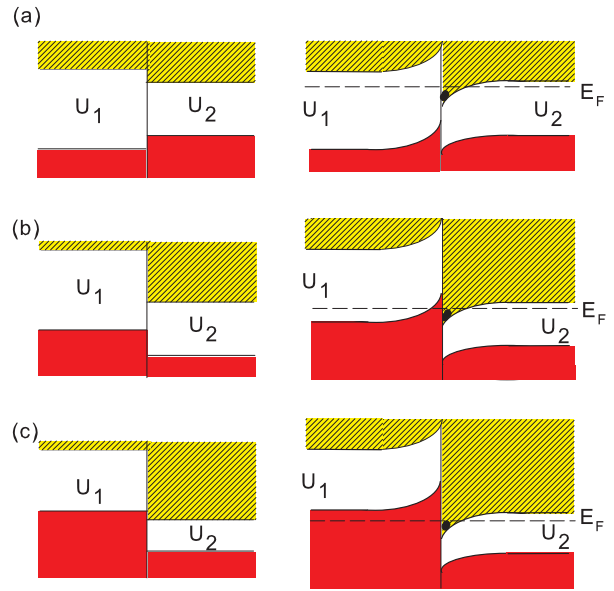}
\caption{\label{fig:one} 
Modulation doping properties of different Mott-insulator Mott-insulator
(MIMI) heterojunction classes.  The figure illustrates the local electronic
spectral function near the interface both before (left) and 
after (right) modulation doping. 
The upper Hubbard band spectral weight (yellow) is 
plotted with shading while the lower Hubbard band (red) is solid.
Mott-Hubbard band bending near the interface is due
to the electrostatic potential induced by the spatial separation of 
dopants and carriers. The discontinuity in 
bands at the interface is determined by atomic scale physics
particular to an individual MIMI heterojunction. In analogy
with semiconductor heterostructure terminology we define the following classes 
of MIMI heterojunctions: (a) Type I: The Hubbard gap
of the smaller gap material is completely inside that of larger gap material. 
Both electrons and holes can then be trapped near the heterojunction, 
depending on doping.(b) Type II: The top of the lower Hubbard band or 
the bottom of the upper Hubbard band of the  
larger-gap material lies inside the Hubbard gap of the smaller-gap material.
Only one sign of carrier can be trapped near the heterojunction in this case.   
(c) Type III: The top of the lower Hubbard band or the bottom of the upper 
Hubbard band of the larger-gap material lies in the opposite band of the smaller-gap material.
In this case charge transfer across the heterojunction occurs even in the absence 
of doping.  Experimental determination of how a particular MIMI heterojunction system
fits in this classification scheme 
is a key element of its characterization.  In this article 
we study only Type I MIMI heterojunctions.}
\end{figure}

Motivated by this recent work we consider in this paper modulation doping 
near an interface between two-different Mott insulators, a MIMI heterojunction.
The model system that we have in mind is sketched in 
Fig. [~\ref{fig:two}].  Most classes of transition metal compounds are either ternary or 
quaternary, with additional {\em spectator} atoms that donate electrons to hybridized transition metal-oxygen
orbitals near the Fermi energy.  These systems can be doped by replacing the spectator atoms
by atoms with a different valence.  Modulation doping of a MIMI heterojuntion is achieved by doping 
the larger gap material at a spectator atom location that is removed from the 
heterojucntion.  The extra electrons then enter the upper Hubbard band of the lower gap 
Mott insulator, creating a two-dimensional doped Mott insulator that is trapped near the heterojunction
by space charge electric fields. The spatial separation between dopants and the carriers
that reside in the upper or lower Hubbard bands should give rise to strongly 
correlated metals that are relatively free from disorder due to chemical 
doping, and are two-dimensional in character.  These systems are illustrated schematically
in Fig.[~\ref{fig:one}].  We study these systems using both HFT and DMFT as in previous studies, 
and also demonstrate that a generalized Thomas-Fermi theory (TFT) can be employed to 
capture key qualitative physics of strongly-correlated heterostructures in a very direct way.  TFT 
yields accurate results for charge-density profiles and for the 
critical doping $\delta_c$ associated with the onset of parallel conduction.
We conclude that both the doping fraction $\delta_{D}$ and the distance between the 
heterojunction and the doping layer play a role in the competition that 
occurs between different magnetically ordered states.  

In the next section we describe the single-band Hubbard model used in this paper to address 
modulation-doped Mott-insulator heterojunction properties.  In Section III
we discuss results obtained for the electronic properties of this model using 
HFT, TFT, and DMFT. In section IV we discuss materials which 
might be suitable for modulation doping of Mott insulator heterojunctions. Finally in Section V
we summarize our findings and speculate on the potential of modulation doped 
Mott insulators. 

\section {Single-band Hubbard Model}

\begin{figure}
\includegraphics{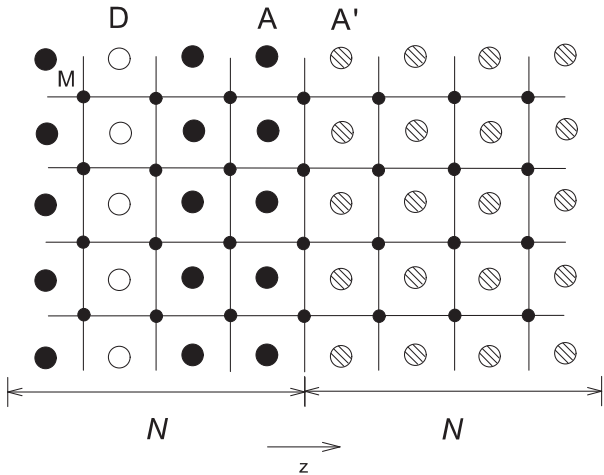}
\caption{\label{fig:two} Schematic representation of the heterostructure
studied in this paper.  We choose the $x-y$ plane as the interface plane and 
$z$ as the layer-by-layer growth direction. The symbol $D$ (white circle) denotes a dopant layer 
containing a fraction $\delta_D$ of dopant atoms with a valence larger than or smaller than the 
$A$ (black circle) and $A'$ (shaded circle) atoms.  Our calculations are performed for a finite thickness
film with $N$ layers of transition metal $M$ atoms (black dots) on
each side of the heterostructure. In the one-band Hubbard model electrons hop between $M$
sites only and are influenced by the space-charge field caused by the spatial 
separation between the dopant atoms and carriers in the upper Hubbard band.}
\end{figure}
The model system we focus on this paper is sketched in Fig. \ref{fig:two}.
The heterostructure is composed of two different $3d^1$ perovskites,
$AMO_3$ and $A'M'O_3$, where both $A$ and $A$' are group III elements and 
$M$ and $M'$ are group IV elements which have nominal $3d^{1}$ electronic structure in this 
structure. Since the total number of electrons per unit cell is odd, compounds of this type must be 
Mott-type when they are insulators, unless translational symmetries are broken.
Modulation doping is achieved by replacing some of the $A$ atoms in the larger gap 
insulator by elements with a different valence.  In this paper we assume electron doping for 
convenience, although the hole doping case is completely equivalent, apart from (important!)
materials specific details which we don't attempt to model in this qualitative study.
If we assume that a fraction $\delta_D$ of the 
$A$ atoms in a single layer of the larger gap material is 
replaced by donor atoms then the sum over all $M$ atom layers of the upper Hubbard band 
fractional occupancy must be $\delta_D$ in order to accommodate the extra electrons.
When modulation doping is successful the added electrons reside in the 
lower gap material, placing them some distance from the ionized donor atoms
and reducing the importance of the chemical disorder normally associated with doping.

For this qualitative study we use a single-band Hubbard model that ignores 
any orbital degeneracies that might be present.  Modulation-doping depends critically on the 
long-range Coulomb interactions so these must be realistically represented in the model. 
Our model Hamiltonian includes hopping, short-range repulsion, and long-range
Coulomb interaction terms $ H=H_t + H_U + H_{Coul}$, where 
\begin{equation}
\begin{array}{l}
\displaystyle
H_t = -t\sum_{<i,j>,\sigma} (d^\dagger_{i\sigma} d_{j\sigma} + h.c.),\\[2mm]
\displaystyle
H_U=\sum_i U(z_i)\; \hat{n}_{i\uparrow}\hat{n}_{i\downarrow},\\[2mm]
\displaystyle
H_{Coul}=\frac{1}{2}\sum_{i\neq j,\sigma,\sigma'}
\frac{e^2 \; \hat{n}_{i\sigma}\hat{n}_{j\sigma'}}{\epsilon \vert\vec{R}_i
-\vec{R}_j\vert} - 
\sum_{i,j,\sigma, I}\frac{Z_I e^2 \; \hat{n}_{i\sigma}}
{\epsilon\vert\vec{R}_i-\vec{R}^I_j\vert},
\end{array}
\end{equation}
$Z_I=1$ for $I=A,A'$ and $(1+\delta_D)$ for $I=D$. 
We do not account for randomness in the dopant layer in this paper.
The index $i$ denotes the position of a transition metal ion ($M$)
so that $\vec{R}_i=a(n_i,m_i,z_i)$ and $\vec{R}^A_i=a(n_i+1/2,m_i+1/2,z_i+1/2)$
respectively in a perovskite unit cell with lattice constant $a$.
For the sake of definiteness, we ignore the possibility of a d-band offset
between the two materials, although these will certainly occur in practice.
Given this assumption, a Type-I MIMI heterojunction will occur whenever the 
Hubbard $U$ parameter is large enough to produce insulating behavior in both materials.
We consider a system with a finite number $2N$ of layers labeled sequentially from
left to right and define 
$U(z_i)=U_1$ for $z_i=1$ to $N$ and $U_2$ for
$z_i= N$ to $2N$ with $U_1>U_2$ so that the larger gap
material is on the left.  We treat the Coulomb part of the interactions in a mean-field Hartree
approximation.  Since Coulomb potentials in the absence of doping are 
implicitly included in the model band Hamiltonian, in evaluating this potential we include only the extra charges
in the dopant layer and charges due to occupancy of lower or upper Hubbard bands.
To be specific, the reference background has charge per atom
equal to $-1$ for each $M$ site and $+1$ for each 
$A$, $A'$, and $D$ site. As a result, the mean-field long-ranged Coulomb
interaction is:
\begin{equation}
H^{eff}_{Coul}=\sum_{i\neq j,\sigma}
\frac{e^2 (\rho_j-1)\hat{n}_{i\sigma}}{\epsilon \vert\vec{R}_i
-\vec{R}_j\vert} -  
\sum_{i,j,\sigma}\frac{\delta_D e^2 \hat{n}_{i\sigma}}{\epsilon\vert\vec{R}_i
-\vec{R}^D_j\vert}
\label{effcou}
\end{equation}
where $\rho_j=\sum_\sigma\langle\hat{n}_{j\sigma}\rangle$ is the electron 
density on site $j$.

\section{Type-I MIMI Heterojunction Electronic Structure} 

\subsection{Hartree-Fock Theory}  

In HFT the strong on-site Coulomb interactions is also treated in a 
mean-field approximation so that 
\begin{equation}
\hat{n}_{i\uparrow}\hat{n}_{i\downarrow} \to \sum_\sigma\langle\hat{n}_{i,-\sigma}\rangle\hat{n}_{i\sigma}.
\label{hft}
\end{equation} 
HFT is equivalent to minimizing the microscopic Hamilton in the 
space of Slater-determinant wavefunctions.
As noted\cite{okamoto3} previously there are typically a number of
self-consistent solutions of the HF equations, corresponding to a number of 
local minima of the Hartree-Fock energy functional.  The various minima usually
are distinguished by different types of magnetic order.  Our philosophy
in examining several different solutions without strong emphasis on their relative HF energies
is that different types of order
will occur near different interfaces but neither the single-band Hubbard model 
nor any of the electronic structure approximations we consider 
(or indeed any known electronic structure approximation) is 
sufficiently reliable to confidently select between them.  Indeed phase transitions
between Mott insulator states with magnetic order and paramagnetic metallic phases,
corresponding to magnetic and non-magnetic extrema of the Hartree-Fock 
energy functional, are often first order.  
(We will however make some conclusions of a more qualitative
nature concerning trends and tendencies related to modulation doping.)     
As explained more fully below, we 
find that the HFT electron density distribution 
near a MIMI heterostructure is sensitive mainly to the relative orientations of 
electron spins on neighbouring metal sites on adjacent layers.  Consequently, we present 
results only for usual bipartite antiferromagnetic (AFM) and ferromagnetic (FM) states,
which in this respect cover the two possibilities.  
These two ordered states are metastable in both undoped and modulation doped 
regimes for the range of parameters we have studied.

\begin{figure}
\includegraphics{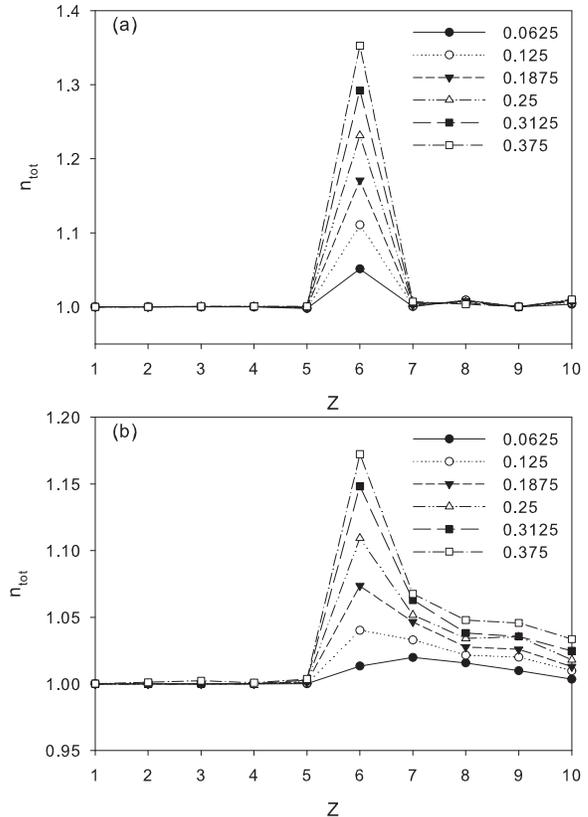}
\caption{\label{fig:three} Electron density distributions from 
HFT for (a) AFM and (b) FM states as a function of $\delta_D$.
The parameters used are $U_1/t=24$, $U_2/t=15$, $U_c=0.8$, and $N=5$.
$z$ is the layer index for $M$ site so that  $U(z)=U_1$ for $z=1 - 5$
and $U_2$ for $z=6 - 10$. The dopant layer is at $z_D=3.5$.}
\end{figure}

The results of our HFT calculations are summarized in Fig. \ref{fig:three}.
We have chosen typical parameters for a one-band Hubbard model of 
perovskite transition metal oxides, taking $U_1/t=24$,
$U_2/t=15$, and $U_c=e^2/\epsilon a t=0.8$.\cite{okamoto1,mizokawa}
We can see from Fig. \ref{fig:three} that the modulation doping
effect occurs for both AFM and FM states, although the details of the electron
density distributions are quite different in the two cases. 
Short-range correlations therefore appear to play a relatively strong role 
in determining the charge distribution near MIMI heterostructures, in contrast to the 
MIBI heterojunction case in which they play\cite{okamoto1} a relatively minor role.
The upper Hubbard band electrons 
are noticeably more confined to the interface in the AFM state case and 
spread further into the smaller-U layer in the FM state case. 
This difference in density-distribution follows from a corresponding 
difference in the compromise between band-energy minimization and 
interaction energy minimization in the two-cases.
The ferromagnetic
state which has all spins parallel maximizes the hopping amplitudes 
between sites, doing a better job of minimizing band energy at a 
cost in interaction energy.  The bandwidth of the mean-field 
quasiparticle states is $\sim t$ for FM
states and $\sim t^2/U$ for AFM states.  Increased doping should favor
FM states over AFM states, at least within HF theory.  From a real-space point 
of view, doping frustrates the staggered moment order of the AFM state more strongly
than it frustrates the FM order because of the nearest-neighbor hopping term $H_t$.
In other words, doping favors the FM state over the AFM state.
  
\begin{figure}
\includegraphics{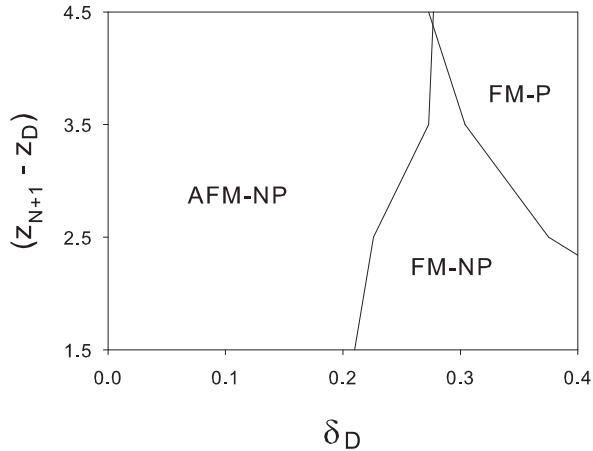}
\caption{\label{fig:four} Ground state ($T=0$) HFT phase diagram {\em vs.}  
doping concentration $\delta_D$ and setback distance $(z_{N+1}-z_D)$. 
AFM-NP denotes the antiferromagnetic state without a parallel conduction channel, 
and FM-NP (FM-P) denotes the ferromagnetic state without (with) a parallel conduction 
channel. Larger setback distances favor antiferromagnetism over ferromagnetism
and strengthen the tendency toward development a of parallel conduction channel.}
\end{figure}

The doped electrons have a strong tendency to accumulate nearly completely in 
one layer in the AFM state case.  Larger setback distances for the dopant layer
should result in larger space-charge fields at the heterojunction and less 
opportunity for electrons to spread out away from the interface, robbing the 
ferromagnetic state of the extra stability that it gains from the third dimension.
We expect therefore that for a given doping level $\delta_D$, antiferromagnetism
will be favored by a larger setback distance for the dopants.  
A larger setback distance also favors the development of parallel
conduction channel. These trends can be seen in the ground-state phase diagram
plotted in Fig.~\ref{fig:four}. 
In summary, modulation doping in MIMI heterostructures may make it 
possible not only to create weakly-disordered low-dimensional strongly correlated 
electron systems, but also to engineer the compromises that occur in these 
systems between different types of magnetic order. 
 
We note in Fig.~\ref{fig:three} that for the FM state, a 
parallel conduction channel starts to appear adjacent to the 
doping layer at $\delta_{D} = 0.375$.  For the parameters
we have chosen modulation doping successfully places the carriers
in a more remote layer up to this doping level. 

\subsection{Thomas-Fermi Theory}

\begin{figure}
\includegraphics{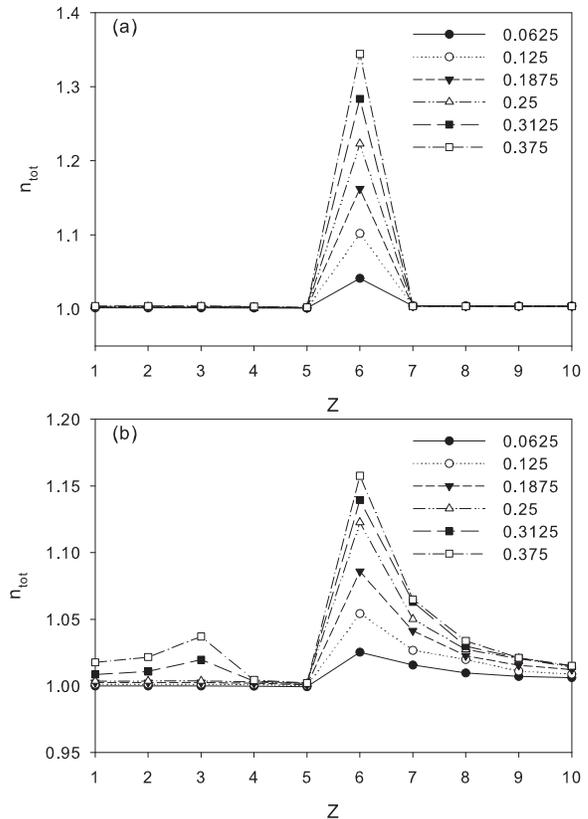}
\caption{\label{fig:five} Total electron density distributions from
TF theory for (a) AFM and (b) FM states with the same parameters as used in Fig
\ref{fig:three}. The results are close to those in
Fig. \ref{fig:three}, except for differences in the critical doping
$\delta_c$ at which modulation doping starts to fail. }
\end{figure}

The HFT results can be understood using a Hubbard-model version of
Thomas-Fermi theory\cite{spruch}. 
The TF equation for this system are:
\begin{equation}
\mu(\rho(z))+v_H(z)=const
\label{tf}
\end{equation}
where $\mu(\rho)$ is the chemical potential at density $\rho$ without 
long-ranged Coulomb interaction and $v_H(z)$ is the electrostatic potential for 
$z$th layer obtained from the charge density by solving the Poisson equation. 
In principle, $\mu(\rho)$ should be
obtained from the exact solution of the three-dimensional Hubbard model.
This input is unfortunately still unavailable. Instead, we can use HFT to obtain 
$\mu(\rho)$.  In this way we have separate versions of the TF 
equations for AFM, FM, and PM states.
As for $v_H(z)$, in the continuum limit each layer
can be approximated by a 2-d uniformly-charged plane so that we have:
\begin{equation}
\frac{v_H(z)}{2\pi U_c}=\delta_D \vert z-z_D\vert-\sum_{z'\neq z}
(\rho(z')-1)\vert z-z'\vert
\end{equation}
where $z_D$ is the layer index for of the dopant layer and $z'$ is summed over 
all electronic layers. Fig. \ref{fig:five} shows results 
calculated using this TFT for the same parameters as used in Fig. \ref{fig:three}.
The total electron density distributions are almost identical to those
obtained from the full microscopic HFT.  We do note that the parallel 
conduction channel in the FM state appears at lower doping in TF 
theory than in the microscopic theory.

The local-density approximation
for correlations implicit in the Thomas-Fermi theory is obviously 
least reliable in judging the relative chemical potentials
for adding carriers to spatially separate layers.  
The discrepancy also occurs partly because the long-ranged Coulomb 
interaction was evaluated using a three-dimensional lattice version
in the microscopic HFT whereas a continuum approximation for the layers
was used in the TFT calculations.  We expect that the TFT model 
is too simplified to determine the critical doping for parallel conduction 
$\delta_c$ accurately for any particular heterojunction, but it may be used
to analyze trends.

\begin{figure}
\includegraphics{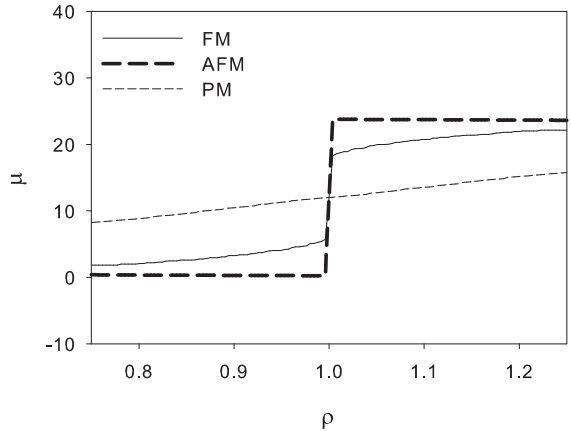}
\caption{\label{fig:six} Chemical potential $\mu$ versus electron density 
$\rho$ for AFM, FM and PM HFT states of the one band Hubbard model with 
$U/t=24$. The jumps at $\rho=1$ for AFM and FM states signal the opening of charge gap.
The exact chemical potential is likely intermediate between the AFM and FM HFT values.}
\end{figure}

The most important consequence of strong local interactions in the Hubbard model is the 
emergence of a jump in the chemical potential when the electron density 
per site crosses from less than one to more than one.  As illustrated 
in Fig.~\ref{fig:six}, the opening of the Hubbard gap is accompanied by 
slower dependence of the chemical potential on density just above and 
just below $\rho(z)=1$, {\em i.e.} by an increase in the thermodynamic
density of states within the Hubbard bands. To capture these features we   
approximate the chemical potential in the upper Hubbard band near $\rho(z)=1$,
by $\mu(\rho(z))\sim E_c(U(z)) + (\rho(z)-1)/{\cal D}(U(z))$,
where $E_c(U)$ is the bottom of the upper Hubbard band, and ${\cal D}(U)$
is the thermodynamic density of states averaged over the energy range of interest near the bottom of the band.
This notation is chosen to emphasize similarities to semiconductor heterojunction physics.
Using this result in each layer we find that 
\begin{equation}
\delta_c \sim \frac{E_c(U_1) - E_c(U_2)}{4\pi U_c (z_{N+1}-z_D)}
+ {\cal O}\left[\left(\frac{1}{{\cal D}(U_{1,2})}\right) \right]
\label{dec}
\end{equation}
where $z_{N+1}$ is the index of the first metal layer on the small U side
of the heterojunction.  This simple and approximate expression emphasizes that 
$\delta_c$ increases with $E_c(U_1)-E_c(U_2)$, decreases with $U_c$ and,
as in the semiconductor case, decreases with the donor layer set-back distance. 
>From HFT we estimate that $E_c(U) \approx U$ for AFM states 
while $E_c(U) \approx U-6t$ for FM states.  In both cases 
$E_c(U_1)-E_c(U_2)$ is approximately
$U_1-U_2$. 
As illustrated in Fig.~\ref{fig:six}, ${\cal D}$ tends to be larger for  
AFM states than for FM states.  The precise form of the thermodynamic density 
of states near the band edge in any particular approximation can only be determined numerically.
This simple expression does not fully capture the difference between 
AFM and FM states, but it does capture some simple but important properties. 
More effective modulation doping will occur materials combinations with larger 
$U$ difference, and smaller $U_c$ ({\em i.e.} larger dielectric constant $\epsilon$)
values. Additionally, because of stronger tendency to confine electrons in one 
layer in AFM state, $\delta_c$ is larger in AFM than in FM states in general.
These features are confirmed by our numerical calculations.

TFT is successful because the dominating energy scales
are the electrostatic energy and the correlation energy arising from local 
correlations.  The ground state electron density distribution is a result of 
competition between these two energy scales, which is accurately captured 
by the TF approximation. 

\subsection{Dynamical Mean-Field Theory}
\begin{figure}
\includegraphics{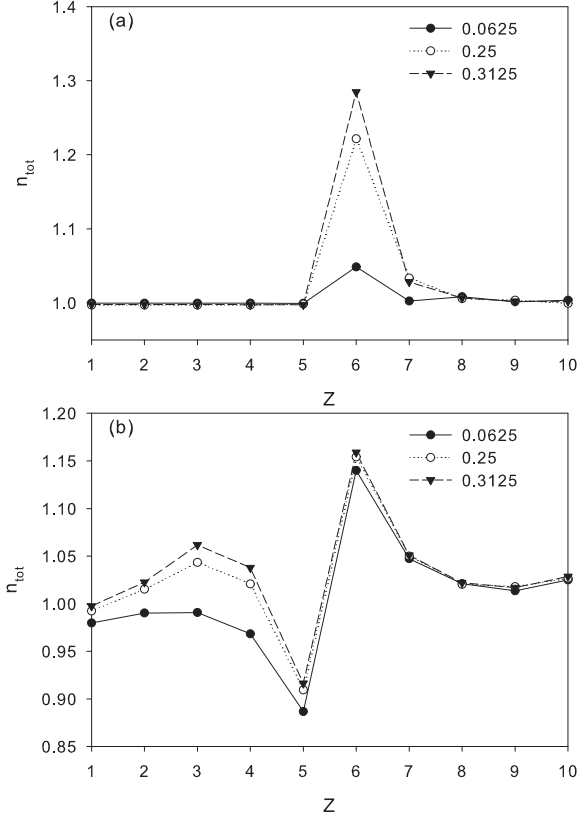}
\caption{\label{fig:seven} Total electron density distributions for the PM state
calculated by (a) DMFT and (b) HFT. The DMFT results exhibit a 
modulation doping effect while those of HFT do not, implying that 
modulation doping near a MIMI heterojunction does not occur without on-site correlations. Note
that Figure (b) can also be reproduced accurately by TF equation 
with the HFT $\mu(\rho)$ of the PM state.}
\end{figure}

HFT provides a particularly poor description of paramagnetic (PM)
strongly correlated states because it is unable to capture 
correlated quantum fluctuations.  In the limit of large $U$ it 
is clear that the thermodynamic properties ($\mu(\rho)$ for example) 
of a paramagnetic state are much more similar to those of 
ordered states than suggested by Hartree-Fock theory.  
To obtain a better description of paramagnetic modulation-doped 
electron systems we appeal to dynamical mean-field-theory 
DMFT\cite{georges1}.  Full DMFT calculations
are however too time consuming, even for our relatively simple model, 
and we therefore employ the two-site method, a 
minimal realization of DMFT that is able to capture the essential physics near a Mott
transition.\cite{potthoff1}  The two-site method has been used 
previously to describe a MIBI heterostructure\cite{okamoto1}. 
Following the general framework of DMFT and the notation used by
Okamoto {\it et al.}\cite{okamoto1}, the electron Green's function for each
in-plane momentum $\vec{k}_\parallel$ can be
written as:
\begin{equation}
G(z,z',\vec{k}_\parallel;\omega)=\left[\omega+\mu-H_t-H^{eff}_{Coul}-
\Sigma(z,z',\omega)\right]^{-1}
\label{green}
\end{equation}
where $H^{eff}_{Coul}$ is given in Eq.~\ref{effcou}. The self energy 
$\Sigma(z,z',\omega)=\delta_{z,z'}\Sigma(z,\omega)$ is obtained by solving a
two-site quantum impurity model 
for each layer and satisfying a set of self-consistency equations\cite{potthoff1,
okamoto1}. We note that the two-site
method predicts that the critical value of $U$ for the metal-insulator transition 
 of a 3-d single band Hubbard model\cite{mp2} to be $U^c \approx 14.7$.
The $U_1$ and $U_2$ values we have chosen are both larger than $U^c$ so that both
perovskites are Mott insulators in the two-site method. 
Fig. \ref{fig:seven} compares the
results from DMFT and HFT for paramagnetic states.  These results demonstrate that modulation doping
is possible without magnetic order in DMFT.  The failure of HFT in this respect is a well 
understood consequence of the importance of on-site
correlation effects for MIMI heterostructure properties, and of the failure of HFT to 
capture these correlations. 

\begin{figure}
\includegraphics{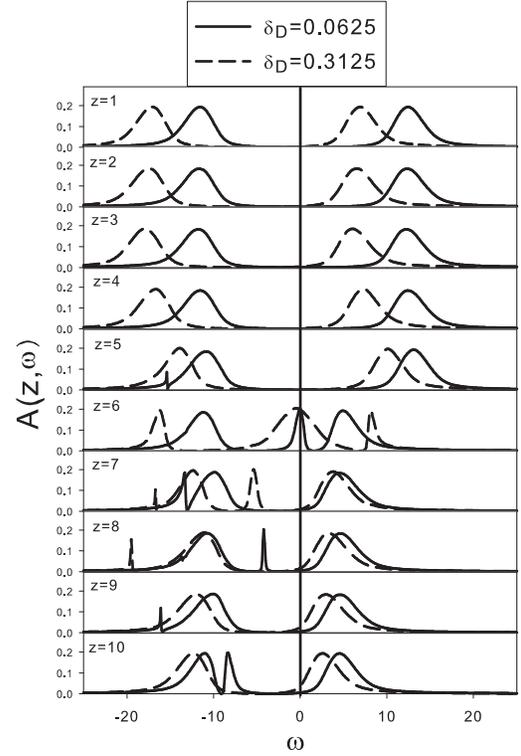}
\caption{\label{fig:eight} Local spectral functions for each layer calculated 
from DMFT for $\delta_D=0.0625$ (solid line) and $0.3125$ (dotted line). 
Only the layer closest to the interface on the smaller-U side ($z=6$) becomes 
metallic upon doping.}
\end{figure}

In Fig. \ref{fig:eight} we plot DMFT layer-dependent electronic spectral functions
\begin{equation}  
A(z,\omega)=-(1/\pi)\int[d\vec{k}_\parallel/(2\pi)^2]\mbox{ Im }
G(z,z,\vec{k}_\parallel;\omega+i0^+).
\end{equation} 
Only the layer closest to the interface on the smaller-U side ($z=6$ in the figure)
develops finite spectral weight near the Fermi surface upon doping.  The appearance of a 
peak in the spectral function near the Fermi energy in layers close to the interface 
is reminiscent of the findings of Okamoto {\em et al.}\cite{okamoto1} for a MIBI heterojunction, who 
refer to this tendency as {\em electronic surface reconstruction}. 
The robustness of this phenomenon beyond the two-site method 
is not certain at present, nevertheless it is intriguing that it 
occurs in two quite different heterojunction systems.
This finding has a natural interpretation in DMFT.  In Eq. \ref{green}, 
$\mu - H^{eff}_{Coul}(z)$ acts like "layer-resolved chemical potential," 
which determines the total electron density in layer $z$. For 
each layer one must solve a separate quasi two-dimensional quantum impurity model whose solution
shows insulating (metallic) behavior for electron density close to (away from) 1. 
The self-consistency equations ensure that all solutions are related.
Consequently, DMFT generally predicts that a layers with electron density away from 
1 (layer 6 in the present calculation) is metallic.

\begin{figure}
\includegraphics{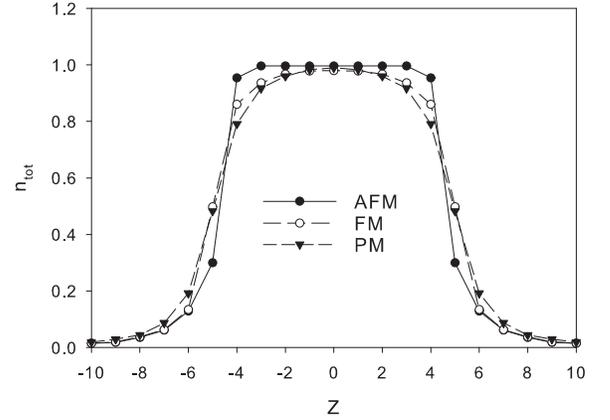}
\caption{\label{fig:nine} Total electron density distributions from
TF theory for AFM, FM, and PM states for the MIBI heterostructure with
$U/t=16$, and $U_c=0.8$. The three-unit-cell-wide crossover is remarkably
reproduced. The similarity in electron density distribution between these
different states demonstrates the relatively weak dependence of MIBI heterojunction
properties on details of the on-site
correlation pointed out by Okamoto and Millis.  The dependence of 
growth direction density distribution on local correlations is stronger
for MIMI heterostructures.}
\end{figure}

\subsection{Thomas-Fermi Theory for MIBI Heterostructures}
Our TFT can accurately reproduce the electron
density distribution near a MIBI heterostructure\cite{okamoto1} calculated
in previous work.  Following the notation of Okamoto and Millis,
the electrostatic potential in the TF relation of Eq. \ref{tf} can be written
as:
\begin{equation}
\frac{v_H(z)}{2\pi U_c}=\sum_{z_{A'}}\vert z-z_{A'}\vert-\sum_{z'\neq z}
\rho(z')\vert z-z'\vert
\end{equation}
where $z_{A'}=\pm 0.5,\pm 1.5,\cdots,\pm 4.5$.
Fig. \ref{fig:nine} shows solutions of the TF equations for
AFM, FM, and PM states which are in accurate agreement with Okamoto's results.
The three-unit-cell wide crossover is clearly seen for each state. The reason
for the weaker dependence on the details of on-site correlations is that many
layers have electron filling smaller than one.  Electrons only fill in
the lower Hubbard band. As a result, the correlation gap between the Hubbard bands, which is sensitive 
to the details of the on-site correlation, does not affect the electronic 
structure very much.

\section {Materials Considerations}
It is still a challenge to determine model parameters
appropriate for particular transition 
metal oxides, although a number of different approaches have been proposed.
\cite{imada,torrance,arima,anisimov}  Although not fully mutually consistent, these
ideas do provide a general picture of how important model parameters vary,
and are a useful rough guide to possible materials combinations that might 
exhibit MIMI modulation doping. 
$RMO_3$ materials ($R$: Rare earth) appear to be an attractive possibility because of their relative simplicity.
It has been shown that YMO$_3$ has stronger electronic correlation than LaMO$_3$
because of the smaller tolerance\cite{imada} factor $f$. Therefore 
YMO$_3$/LaMO$_3$ heterostructure appear to be a good candidate for
realizing modulation doping.  In particular LaTiO$_3$ and YTiO$_3$ are both Mott
insulators with distorted perovskite
structures (GdFeO$_3$ type) having gaps $\approx 0.2\mbox{ eV}$ and 
$\approx 1\mbox{ eV}$ respectively\cite{okimoto}. They might be used to 
realize a modulation doped heterostructure if YTiO$_3$ could be doped.
We emphasize that at present we do not know how the $t_{2g}$ d-bands are 
aligned at a heterojunction between these two materials.  Indeed, aside
from the band-lineup issue, it is importnat to recognize that the simple model
considered in this paper is not sufficiently rich to capture all aspects of the interface physics 
that can be relavant to modulation doping and to magnetic order in the interface layer.
For example orbital degeneracy plays a key role in the magnetic state of 
bulk YTiO$_3$ and in all liklihood would also play a role in determining 
the magnetic state of any two-dimensional matallic layer at the interface.

Although our one-band model is intended to qualitatively describe $3d^1$ 
systems with cubic perovskite structure, 
some of our results should be generalizable. As emphasized above, the modulation 
doping effect is a consequence of Coulomb space-charge fields and on-site correlations. 
We therefore do not expect that perfect affinity to the ideal perovskite structure 
to be of key importance. Other $RMO_3$ type Mott insulators with $R$=rare earth or
alkaline earth and $M=$Mn,Cr might also be good candidates, although there will certainly 
be additional complications because of the larger $d$-valence that 
are not addressed at all in this work.
Building up more realistic models for potential building block materials
is an important challenge for theory. 

\section {Summary and Conclusions}
In this paper, we have presented some theoretical considerations related 
to modulation doping near heterojunctions between two different Mott
insulators, combining insights from Hartree-Fock Theory, Thomas-Fermi 
Theory, and Dynamical Mean-Field-Theory approaches.  Using typical
parameters within a simple single-band Hubbard model, we predict that 
modulation doping is possible with doping layers set back from the 
heterojunction by several lattice constants.  Modulation doping can
be used to create 2-d strongly correlated electron system with 
weak disorder and controllable densities.  These systems could prove to 
be an interesting platform for systematic studies of strongly correlated system.
Indeed, we find that the magnetic phase diagram can be altered not only
by the dopant density $\delta_{D}$ but independently by the 
dopant layer setback distance. Unlike the case of MIBI heterostructures in which 
the Coulomb field dominates, the growth direction electron density distribution in 
MIMI heterojunction systems also depends strongly on the character of in-plane ordering.
In the AFM state electrons are more confined to the interface while in the FM state electrons
spread further away from the heterojunction into the smaller gap material. 
The results of HFT can be reproduced remarkably well by 
Thomas-Fermi theory, indicating that the electrostatic energy and the 
correlation energy resulting from local fluctuations dominate the 
physics of the heterostructure.  From the TF equation, we estimate the critical doping 
$\delta_c$ at which modulation doping starts to fail. 
DMFT calculations show that modulation doping can occur without magnetic order 
, and that it requires only the on-site correlations 
that lead to the Mott-Hubbard gap.  Layer-dependent 
spectral functions calculated using DMFT indicate that only the interface layer is metallic,
reminiscent to the earlier findings of Okamoto {\em et al.}\cite{okamoto1}.

Doped Mott insulators typically appear to have exotic properties when the doping
is small and more conventional properties when the doping is large and the 
total band filling is well away from one, the value at which local correlations 
have maximum importance.  In the case of the extremely heavily studied cuprate 
systems, for example, this crossover is interrupted, by high-temperature 
superconductivity.  It is interesting to consider whether or not the two-dimensional electron
systems considered in this paper are Fermi liquids.  Whereas bulk doping often
leads eventually to a first order transitions between a doped Mott insulator and a 
relatively conventional metal, modulation doping in a single or several layers 
may make it possible to realize high-density, low-disorder, two-dimensional exotic metals which 
carry reflect the heritage of the three-dimensional Mott insulators from which they emerge.
Since the very existence of these two-dimensional electronic systems 
depends on gaps that are entirely due to electron-electron interactions, it is 
clear that they cannot be adiabatically connected to non-interacting electron states.
On the other hand, in the HFT description the doped state {\em is} a Fermi liquid with well defined 
quasiparticles.  This approximation neglects quantum fluctuations of the 
magnetic state however, and its predictions for quasiparticle properties may
not be reliable.   

We have also speculated briefly on materials combinations that might be 
attractive to realize the physics discussed in this paper. 
Predictions of the phase diagram for particular materials 
combinations will require much more detailed modeling, 
and may be assisted by insights from experiment as well
as from {\em ab initio} electronic structure\cite{Noguera2004,Sawatzky} calculations.
A detailed description would require many realistic features 
of perovskite materials to be addressed, for instance lattice distortions\cite{goodenough,sawatzkypreprint}
which may vary with proximity to the interface, and related orbital degeneracy issues. 
Progress will require progress in materials growth and characterization and 
interplay with {\em ab initio} and phenomenological modelling. 

\section{Acknowledgments} 
This work was supported by the Welch Foundation.  The authors acknowledge helpful discussions
with Bernhard Keimer, Karin Rabe, George Sawatzky, David Singh, and John Goodenough. W.-C. Lee would like to thank Satoshi Okamoto for sharing precious experiences in DMFT.

\end{document}